\documentclass[11pt,twoside]{article}
\usepackage{asp2010}
\usepackage{graphicx}

\resetcounters

\newcommand{\teff}{$T_{\mathrm{eff}}$}
\newcommand{\muhz}{$\mu$Hz}
\newcommand{\numax}{$\nu_{\mathrm{max}}$}
\newcommand{\dnu}{$\Delta\nu$}
\newcommand{\astec}{\textsc{ASTEC}}
\newcommand{\adipls}{\textsc{ADIPLS}}

\begin{document}

\title{Non-radial modes in cool stars}
\author{Dennis~Stello$^1$
\affil{$^1$Sydney Institute for Astronomy (SIfA), School of Physics, University of Sydney, NSW 2006, Australia}}

\begin{abstract}
In cool stars that oscillate like the Sun, non-radial modes become mixed
as the stars evolve.  The mixing is caused by the coupling between g-modes
in the stellar core and p-modes in the envelope, which results in
distinctly different and more complex frequency spectra for subgiants and
red giants than seen in
main sequence stars.  Using a new version of the `scaled' \'echelle diagram, I 
illustrate how the frequencies of non-radial modes evolve during the
evolution from the main sequence to the red giant branch, and
I show how they depend on stellar mass and metallicity.  Then, with focus on 
the dipole ($l=1$) modes, which show the strongest effects from mixing, I
present a toy model to fit, and hence identify, those modes in a large series
red giant models.  
\end{abstract}

\section{Introduction}
Stellar models predict that non-radial oscillation modes in stars like the
Sun will become of a mixed character as the stars evolve off the main
sequence \citep{Osaki75}.  The mixing, which is caused by the coupling of
two modes (a 
g-mode in the stellar core and a p-mode in the envelope) will result in
shifts of the mode frequencies - known as mode bumping - as the two modes
undergo so-called avoided crossings. 
As a result, mixed modes do not follow the usual asymptotic relation of
regular spaced high radial order p-mode frequencies, $\nu$, as expressed by 
\citep[e.g.~][]{Vandakurov68,Tassoul80,Dalsgaard11}   
\begin{equation}
 \nu_{nl}=\Delta\nu(n+l/2+\epsilon)-\delta\nu_{0l}
\label{asymptot}
\end{equation}
where the small frequency separation 
\begin{equation}
 \delta\nu_{0l}\propto(l+1/2)^2\frac{\Delta\nu}{\nu_{nl}}\,.
\label{smallsep}
\end{equation}
Here $n$ is the radial order, $l$ is the degree, $\epsilon$ is a dimensionless
offset, and \dnu~is the separation between over tone modes of the same
degree. 
When visualized in the \'echelle diagram, where we plot mode frequency
versus the frequency modulo the large frequency separation, \dnu, mixed
modes can show strong departures from the almost vertical ridges of the
non-mixed modes, which I will discuss next.

\section{The scaled \'echelle diagram}
A useful diagram to compare mode frequencies of many stars (or models) is
the scaled \'echelle diagram introduced by \citet{BeddingKjeldsen10}.   
Their Method 2 sought to make the ridges of the radial modes coincide for
different, though still similar, stars by fine tuning \dnu; in essence removing any differences
in $\epsilon$ by allowing the ridges to depart slightly from vertical.
This provides a clear picture of how the modes of different degree are
positioned relative to one another in a comparative analysis.  Here I will
use a slightly modified version of this method to show the evolution of a
large series of models ranging from the main sequence to the red giant
branch (RGB).  For each model, \dnu~is derived from the four
radial modes closest to the frequency of maximum power, \numax~(scaled from
the solar value according to \citet{KjeldsenBedding95}), and the
variation in $\epsilon$ from one model to the next is removed by shifting
the frequencies to a common value of $\epsilon=1.2$; hence maintaining
strictly vertical ridges of the radial modes.  This could therefore be
appropriately called a shifted \'echelle diagram. 

\begin{figure}
\center
\includegraphics[width=8.cm]{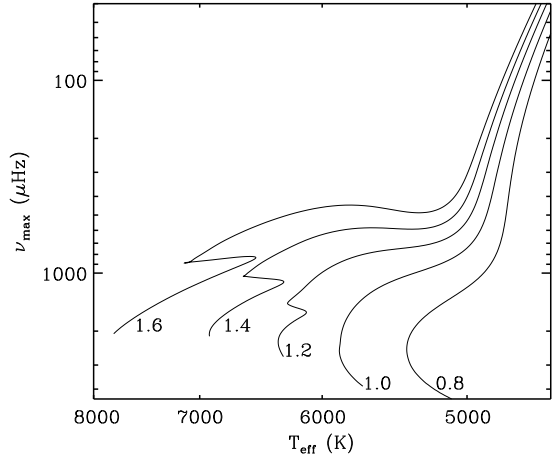}
\caption{Asteroseismic Hertzsprung-Russell diagram with \numax~as
  luminosity proxy.  Tracks are for $Z=0.017$ and their masses in solar
  units are shown at the ZAMS. 
\label{hrd}} 
\end{figure}

The stellar models were derived with \astec~\citep{DalsgaardAstec08} and adiabatic
frequencies were calculated using the \adipls~code
\citep{DalsgaardAdipls08}. The models range in mass from 0.8$M_\odot$, to
1.6$M_\odot$ in steps of 0.2$M_\odot$, and in $Z$ from 0.011 to 0.028 in
steps of 0.1 dex in $\log Z/X$ (with $X=0.7$ fixed).  The tracks were evolved from the
zero-age-main-sequence (ZAMS) towards, but not quite reaching, 
the tip of the RGB.  Details on the model physics, can be
found in \citet{Stello09} and references therein.  The tracks for $Z=0.017$ 
(approximately solar metallicity) are shown in Figure~\ref{hrd}, where I
have used \numax~ as the proxy for
luminosity to make it more useful in comparison with the following figures.

\begin{figure}
\includegraphics[width=13.cm]{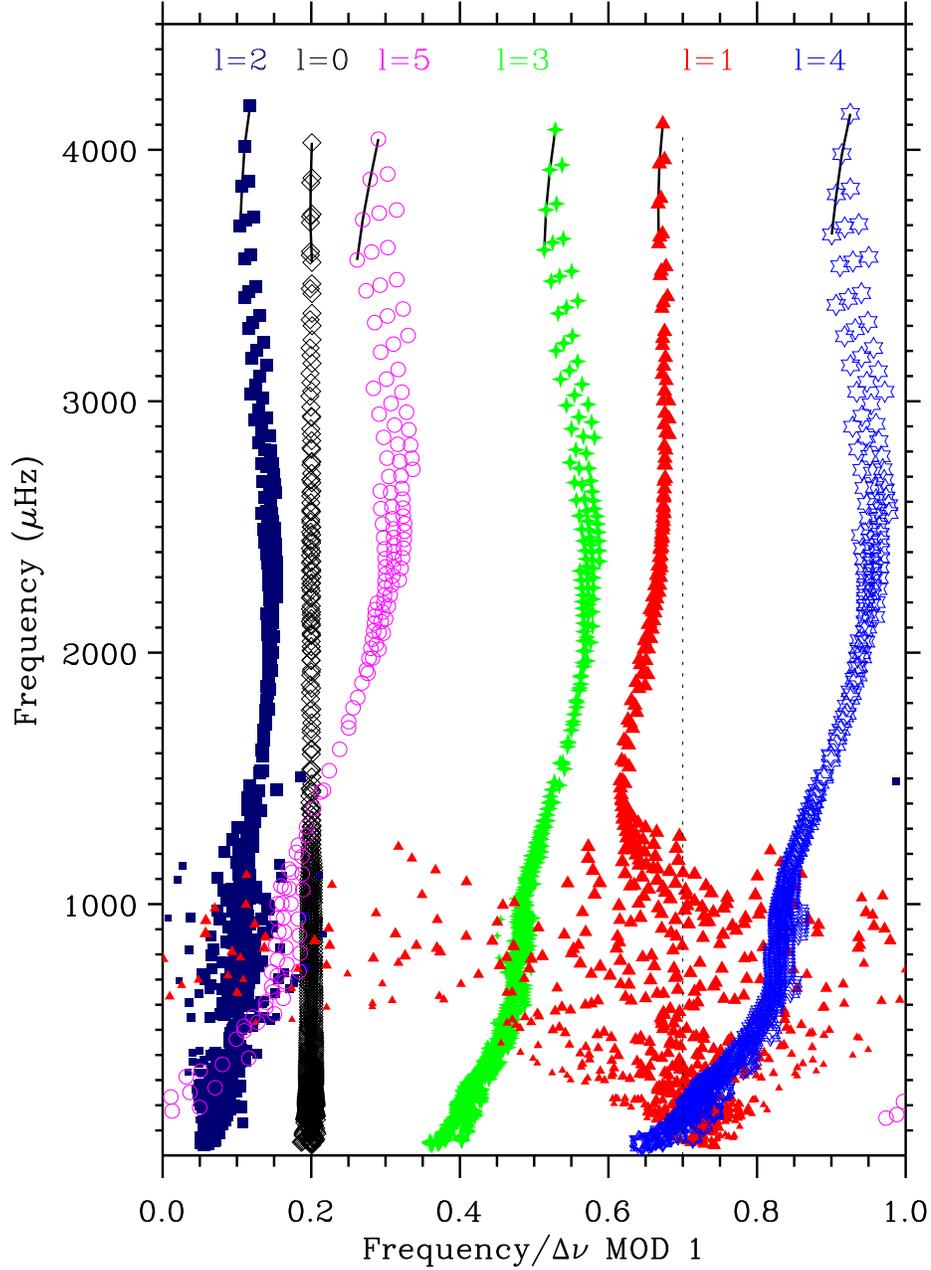}
\caption{\'Echelle diagram for series of models along a 1.0M$_\odot$ track
  ($Z=0.017$, [Fe/H$]\simeq 0.0$; see Figure~\ref{hrd}).  For each model the four modes closest
  to \numax~are plotted after they were shifted to a common value of
  $\epsilon=1.2$. The dotted line shows the halfway point between adjacent
  radial modes. \label{f1}} 
\end{figure} 

In Figure~\ref{f1}, I show an \'echelle diagram for a series of models 
along the one solar mass track, where four orders are plotted for each
model.  The symbol sizes are scaled according to the reciprocal of the
square root of the mode
inertia, normalized to the radial modes, but set to zero below a
certain threshold for clarity.   
Several features are apparent in this diagram.
\begin{enumerate}

\item We see an increasing tilt of
  the non-radial ridges with increasing $l$.  To illustrate this,
  I have connected modes of the same degree by lines for the ZAMS model.
  The tilt can be explained by the last term in the asymptotic
  relation, $\delta\nu_{0l}$ (Eq.~\ref{smallsep}), where a higher $l$
  implies a higher frequency dependence, and hence a stronger tilt.  

\item The position of the non-radial ridges, and hence the small separations,
  changes quite significantly with evolution.  
  After moderate changes during the main sequence, there seems to be a
  slight pause during the subgiant phase where the stellar density is not
  changing by much, before more dramatic changes
  occur during the ascent of the RGB.  
  That the change is larger for increasing mode degree is explained by the
  same asymptotic term as for the above tilt.  
  Interestingly, this evolution of the ridge positions shows some   
  similarities to the variation seen in $\epsilon$ as a function of
  \numax~\citep{White10}.

\item At a certain point during the evolution, corresponding roughly to the
  transition between the main sequence and the subgiant phase, mixed modes
  start to appear, most clearly for $l=1$ and $l=2$, leading to modes
  spreading across the entire echelle diagram ($l=1$ in particular). 

\item After the subgiant phase the mixed modes become more confined as the
  models evolve up the RGB. Ultimately, near the tip of the
  RGB, the coupling between p- and g-modes become so weak
  (efficient mode trapping) that we expect to see only one mode per degree
  per radial order, as in the main sequence phase \citep{Dupret09}.  

\item As first observed by \citet{Bedding10} and shown for stellar models by
  \citet{Montalban10}, the small separation, $\delta\nu_{10}$, between the
  the $l=1$ modes and the point halfway between consecutive radial modes
  (dotted line) becomes negative for the red giant models.  

\end{enumerate}

Note that the less dense $l=5$ ridge arise simply because these modes were
not derived for all models along the evolution track at the subgiant and
red giant phases.

\subsection{Changing mass and metallicity}
\begin{figure}
\includegraphics[width=13.cm]{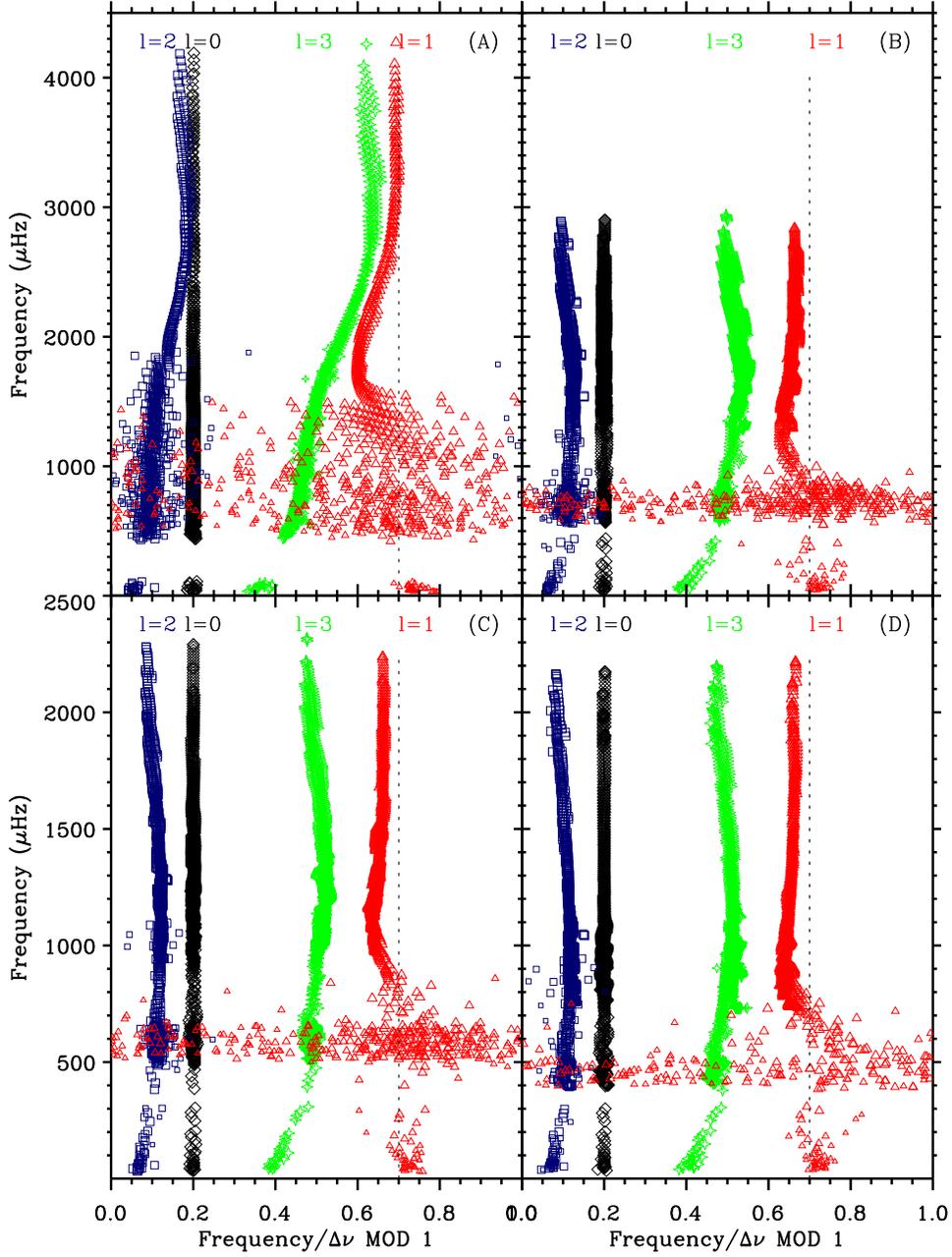}
\caption{\'Echelle diagram for series of models along the A: 0.8M$_\odot$,
  B: 1.2M$_\odot$, C: 1.4M$_\odot$, and D: 1.6M$_\odot$ tracks
  ($Z=0.017$, see Figure~\ref{hrd}). The y-axes are rescaled for the lower
  panels for improved visibility. 
\label{f2}} 
\end{figure}

\begin{figure}
\includegraphics[width=13.cm]{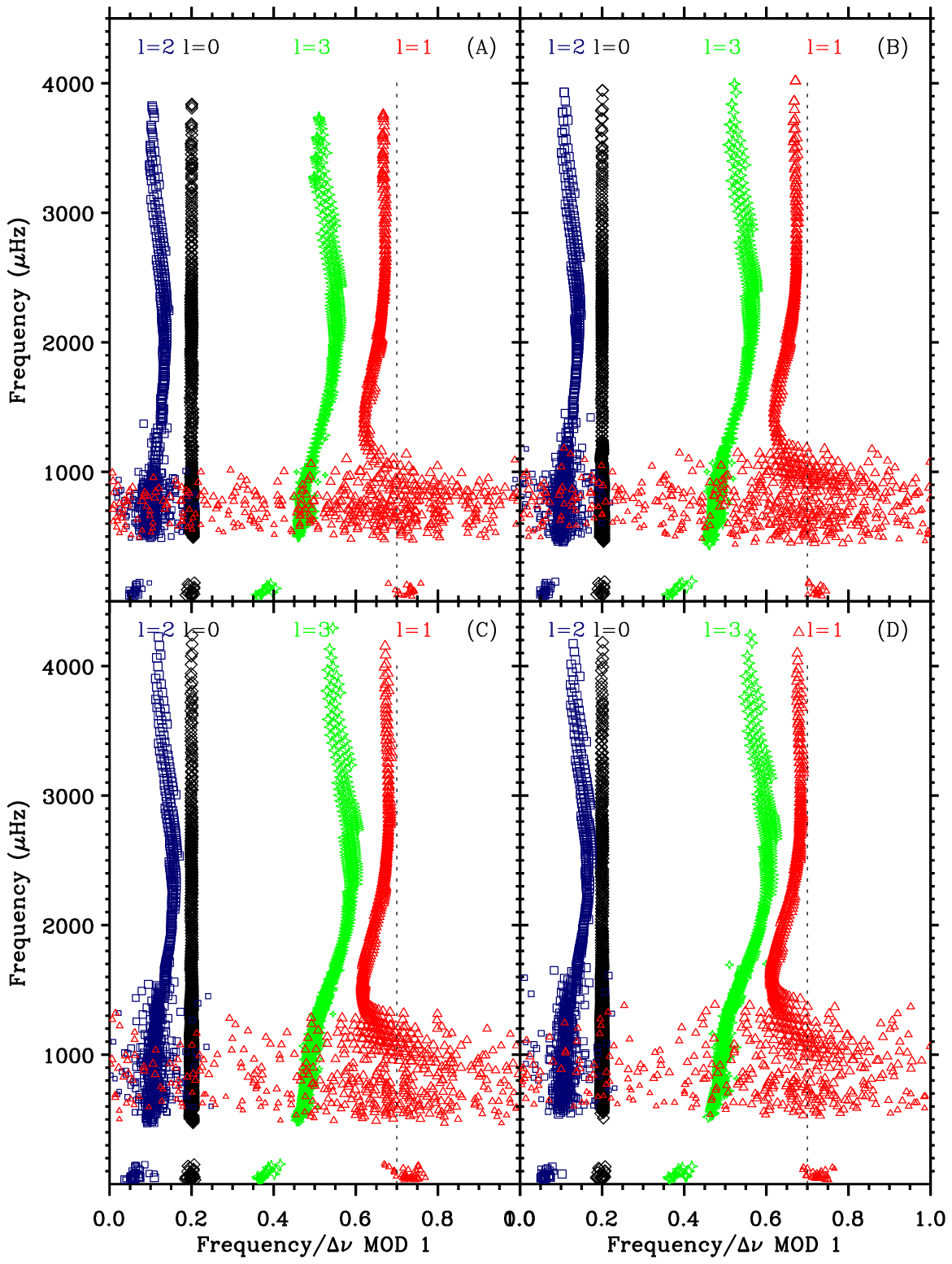}
\caption{\'Echelle diagram for series of models along 1.0M$_\odot$ tracks
  of A: $Z=0.011$ ([Fe/H$]\simeq -0.19$), B: $Z=0.014$ ([Fe/H$]\simeq -0.08$),
  C: $Z=0.022$ ([Fe/H$]\simeq 0.11$), and D: $Z=0.028$ ([Fe/H$]\simeq 0.22$).  
\label{f3}} 
\end{figure}

To illustrate how the different features seen in the \'echelle diagram
depend on stellar mass and metallicity, I show examples for $l\leqq3$ of
models with a range of masses and metallicities (Figures~\ref{f2},
\ref{f3}).  Due to the high computational requirements for deriving mode
frequencies of evolved stellar models there is a lack of frequencies
in the model grid, which appears near the base of the RGB
(300\muhz~$\lesssim$ \numax~$\lesssim 500$\muhz).  Fortunately, the highly evolved
red giant models at even lower \numax~enables 'interpolation by eye' to
bridge the gap. 


It is evident from Figure~\ref{f2} that the ridge positions change
significantly as a function 
of mass, with lower masses showing smaller separations between the odd
ridges ($\delta\nu_{02}$) and even ridges ($\delta\nu_{13}$), respectively.
The higher mass models, which exhibit the 
smallest density changes during the subgiant phase, show more static
ridge positions during this phase in agreement with Eq.~\ref{smallsep} and
the observation made in Figure~\ref{f1} (point 2).  We further see a hint that the
$l=2$ modes spread out more for the least massive models, indicative of
stronger mode coupling. 

If we turn our attention to the effect from metallicity (Figure~\ref{f3}), we
see that an 
increase in $Z$ results in a decrease in the separations between the odd
ridges and even ridges, respectively.  This is in qualitative agreement
with the effect we saw on mass, namely that the cooler the model the smaller
the separations $\delta\nu_{02}$ and $\delta\nu_{13}$ become.

\subsection{Mixed dipole modes in red giants}
In the following I will focus on the lower part of Figure~\ref{f1} where we see
increased complexity in the $l=1$ ridge due to mode mixing.  In particular
I will address how to identify the $l=1$ modes in frequency spectra of red 
giants. 

\begin{figure}
\includegraphics[width=13.cm]{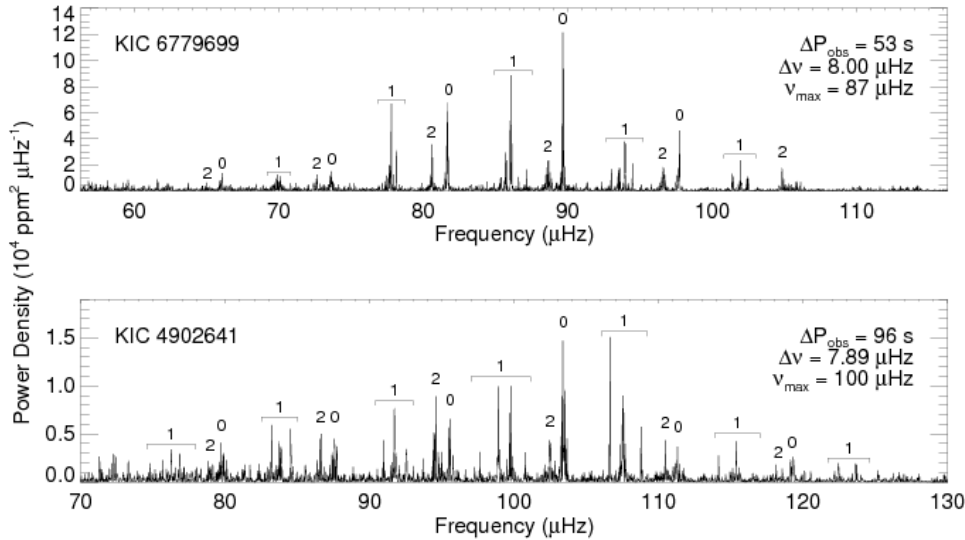}
\caption{Frequency spectra of an RGB star and a red clump star (taken from
  \citealt{Bedding11}). Mode degree is indicated.   
\label{f4}} 
\end{figure}

The complicated frequency spectra of $l=1$ modes in subgiants can in many
cases be successfully fitted, resulting in correct identification of the mode
degree (\citealt{DeheuvelsMichel10}, Benomar et al. in prep.).  However,
can something similar be done for red giant stars?   
To motivate this question, I show the frequency spectra of two red giants
in Figure~\ref{f4}.  
The weaker mode coupling in red giants imply we do not see the $l=1$ modes 
appear everywhere, as in subgiants (Figure~\ref{f1}), but rather within well confined
clusters, one for each radial order (centered at the frequency of the
'unperturbed' p-mode in the hypothetical case where there would be no
coupling with the g-modes 
in the core).  On the one hand, this makes the spectra more simple as we
avoid a very dense spectrum of $l=1$ modes that overlaps with all the other
modes.  On the other hand, the slightly complicated frequency separations
of these modes \citep{Beck11,Bedding11} makes it difficult to estimate 
how many modes are in fact 'missing in the gaps' between each cluster of
modes, and hence hinders a straightforward determination of the
frequencies of the undetected modes.  In practise this means that it can
become difficult to make judgment on whether a peak in the observed
frequency spectrum belongs to the sequences of $l=1$ modes or whether
it is due to a mode of a different degree, or simply noise.   

Fortunately, for stellar models we have frequencies for all modes
available, even those in the gaps that are not detectable in reality.
This can help us develop ways to analyse the complex frequency spectra
observed in red giants.  
As a first step towards fitting observed $l=1$ modes I will show
some preliminary results for what one obtains when fitting a simple toy
model to the stellar model frequencies.  The aim of this exercises is to 
identify relations between on the one hand the parameters that describe the
toy model and one the other hand the quantities that we are able to detect
in the observed frequency spectra -- ultimately enabling us to make informed
decisions and on how to interpret the observations.

\begin{figure}
\includegraphics[width=13.cm]{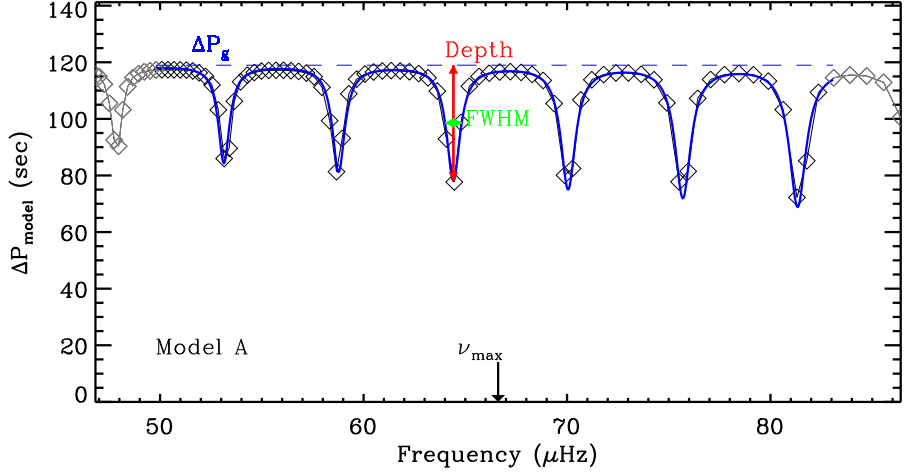}
\caption{Period spacing versus frequency for dipole modes (diamonds) of
  Model A, which has \teff~$=4745\,$K and $L/\mathrm{L}_\odot=56$ (marked
  in Figure~\ref{f6}b). Toy model fit is shown by thick solid blue line.
  Resulting fitting parameters, $\Delta P_g$, FWHM, and Depth are indicated
  (see text for details).
\label{f5}} 
\end{figure} 
Figure~\ref{f5} shows the period spacing of the $l=1$ modes from a model
located on the RGB.  The period spacing is approximately constant as
expected from asymptotic theory of pure g-modes \citep[e.g.~][and references
  therein]{Dalsgaard11}.  The
departure from the constant spacing, seen as regular dips, 
are caused by the coupling between the g- and p-modes, which shifts
the frequencies.  The dips are separated by \dnu~-- the separation between
the unperturbed p-modes. 
The most weakly coupled modes are almost pure g-modes, hence showing
near equal period spacings. However, being pure g-mode means they are
basically contained in the core, and hence not observable.  The more
strongly coupled modes, which show the largest departure from equal period
spacing, are more p-mode like, and have therefore significant amplitudes at
the surface and are the ones we detect as small clusters. 

\begin{figure}
\includegraphics[width=13.cm]{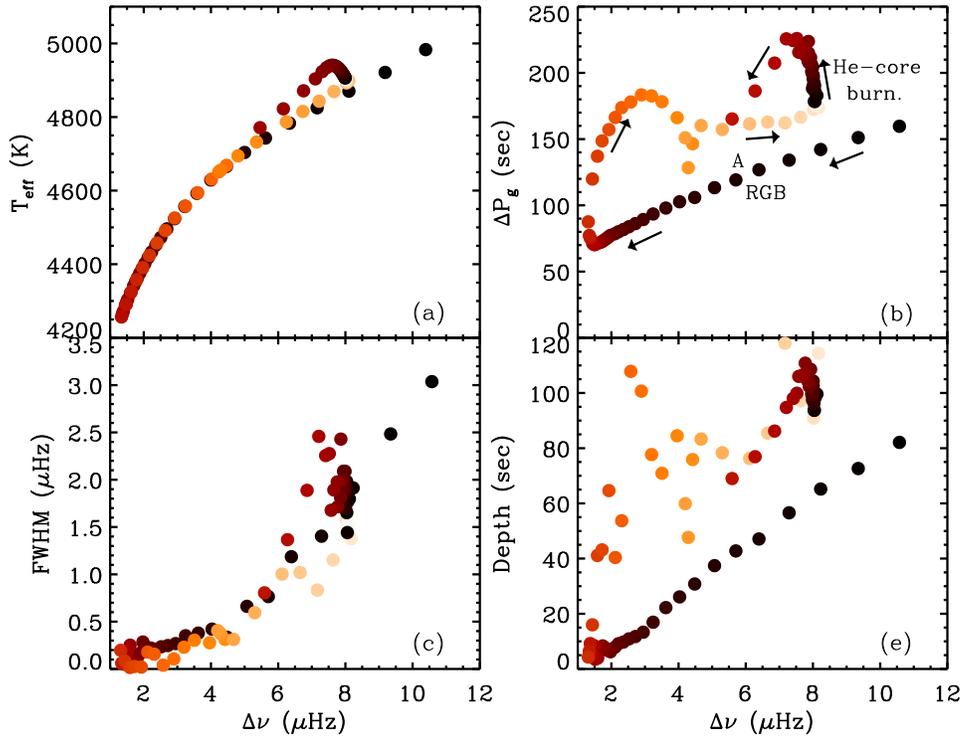}
\caption{(a) Evolution track of 2.4M$_\odot$ track. (b)-(d) Results of
  fitting the toy model (Figure~\ref{f5} thick blue curve) to each model along the
  track. Plotted values of FWHM and Depth are evaluated at \numax. 
\label{f6}} 
\end{figure} 
I have simultaneously fitted a series of Lorentzian profiles, one for each
dip to a series of stellar model frequencies of dipole modes.  The free
parameters in the fit are $\Delta P_g$ (the true g-mode spacing unperturbed
by mixing with p-modes), FWHM (the width of the Lorentzian),
$\delta$(FWHM)/$\delta f$ (the change in the width as a function of
frequency), Depth (the depth of the dip), $\delta$(Depth)/$\delta f$ (the
change in the depth as a function of frequency), and the frequency of the
centre of each Lorentzian profile, assumed to be strictly equally spaced.
Figure~\ref{f6}a shows a 2.4M$_\odot$ track evolving from  
the base of the RGB 
to the base of the asymptotic giant branch (like a transposed Hertzsprung
Russell diagram), and Figures~\ref{f6}b-d show the
result of fitting the toy model to the stellar models along that track.  An 
example of a fit is shown by the blue curve in Figure~\ref{f5}. 
Interestingly, there seems to be a strong relation between $\Delta P_g$ and
Depth, but Depth is not simply $\frac{1}{2}\Delta P_g$ as suggested by
\citet{Bedding11}.  Although FWHM and Depth are both related to the coupling
strength between the g- and p-modes, only Depth separates the RGB from the
red clump, while FWHM is largely a function of \teff.   
When repeated for stellar models of different masses, the relations between 
the toy model parameters and the seismic observables change.  Quantifying
these relations will help create a formalism to fit mixed mode frequencies
from observations. 

\section{Summary and future work}
I have shown that a new version of the `scaled' \'echelle diagram can be
useful to compare an ensemble of stellar models, and in particular to show
the evolution of pulsation frequencies from the main sequence to the RGB as
well as how 
the frequencies depend on mass and metallicity.  
For red giants we see that the $l=1$ modes become more confined as small
clusters of modes.  While this in some sense makes the frequency spectra
simpler, it also brings new challenges by making it harder to determine
the frequencies of all the modes between each cluster.
I briefly demonstrated that fitting a simple toy model to stellar model
frequencies can aid in establishing relations between seismic observables
and the toy model parameters. When fully developed this could lead to
a robust way to fit the observed $l=1$ frequencies and hence to determine
the characteristics of their complex frequency separations.

\acknowledgements I would like to thank Hiromoto Shibahashi for support to
attend this conference, and all participants who despite the unusual
circumstances took this as an experience...Kampai!.  I would also like to
thank J\o rgen Christensen-Dalsgaard for comments on manuscript.


\end{document}